\documentclass[aps,prl,amsmath,amsfonts,superscriptaddress,twocolumn]{revtex4}
\usepackage{epsfig,psfig,graphicx}

\newcommand{\beq}{\begin{equation}}
\newcommand{\enq}{\end{equation}}

\begin{document}

\title{Quantum phase transition in an atomic Bose gas with a Feshbach resonance}

\author{M.W.J. Romans}

\affiliation{Institute for Theoretical Physics, Utrecht
University, Leuvenlaan 4, 3584 CE Utrecht, The Netherlands}

\author{R.A. Duine}

\affiliation{Institute for Theoretical Physics, Utrecht
University, Leuvenlaan 4, 3584 CE Utrecht, The Netherlands}

\author{Subir Sachdev}

\affiliation{Department of Physics, Yale University, P.O. Box
208120, New Haven, Connecticut 06520-8120, USA}

\author{H.T.C. Stoof}

\affiliation{Institute for Theoretical Physics, Utrecht
University, Leuvenlaan 4, 3584 CE Utrecht, The Netherlands}

\begin{abstract}
We show that in an atomic Bose gas near a Feshbach resonance a
quantum phase transition occurs between a phase with only a
molecular Bose-Einstein condensate and a phase with both an atomic
and a molecular Bose-Einstein condensate. We show that the
transition is characterized by an Ising order parameter. We also
determine the phase diagram of the gas as a function of magnetic
field and temperature: the quantum critical point extends into a
line of finite temperature Ising transitions.
\end{abstract}


\maketitle

{\it Introduction.} --- One of the most important recent
developments in the field of ultracold atomic gases is the
application of Feshbach resonances. A Feshbach resonance in the
scattering amplitude of two atoms occurs when the total energy of
the atoms is close to the energy of a molecular state that is
weakly coupled to the atomic continuum
\cite{stwalley1976,tiesinga1993}. In the alkali gases of interest,
this coupling is provided by the exchange interaction, and
consequently the magnetic moments of the two atoms and the
molecule differ substantially. As a result the energy difference
between the molecule and the threshold of the two-atom continuum,
known as the detuning $\delta$, can be experimentally tuned by
means of a magnetic field. Moreover, by sweeping the magnetic
field from positive to negative detuning through the Feshbach
resonance, it is actually possible to form molecules in the
atomic gas
\cite{regal2003b,strecker2003,bourdel2003,grimm2003,ketterle}.

Very recently, it has even been possible to create a Bose-Einstein
condensate (BEC) of molecules in an atomic Fermi gas with a
Feshbach resonance \cite{jochim2003,greiner2003,zwierlein2003}.
This achievement is of particular importance because it offers the
opportunity for observing a Bardeen-Cooper-Schrieffer (BCS)
transition in a dilute atomic gas \cite{stoof1996a,stoof1996b}.
The idea here is to start from a pure molecular condensate and
then sweeping the magnetic field back to positive detuning to
create a much colder atomic Fermi gas than could have been
achieved without the intermediate step of forming a molecular
Bose-Einstein condensate. It is important to realize that this
idea strongly hinges on the fact that a smooth BEC-BCS crossover
exists in this system as one changes $\delta$
\cite{ohashi2002a,ohashi2002b,milstein2002}. In this Letter we
show that analogous experiments varying $\delta$ in an atomic Bose
lead to a true phase transition rather than a crossover
\cite{nozieres}. The reason for, and the nature of, this phase
transition can be understood as follows.

The argument contains two important ingredients. The first
ingredient is that the coupling between the atoms and molecules in
the gas is provided by an interaction energy that is proportional
to $\int d{\bf x}~\left( \psi_{\rm m}^{\dagger}({\bf x}) \psi_{\rm
a}({\bf x}) \psi_{\rm a}({\bf x}) + \psi_{\rm a}^{\dagger}({\bf
x}) \psi_{\rm a}^{\dagger}({\bf x})\psi_{\rm m}({\bf x}) \right)$,
where $\psi_{\rm a}({\bf x})$ and $\psi_{\rm m}({\bf x})$
annihilate an atom and a molecule at position ${\bf x}$,
respectively \cite{drummond1998,timmermans1999b,RembertReview}.
Such a coupling implies that if the gas contains an atomic
Bose-Einstein condensate, and therefore has a nonzero value of
$\langle \psi_{\rm a}({\bf x}) \rangle$, the gas must necessarily
also contain a molecular Bose-Einstein condensate and have a
nonzero value of $\langle \psi_{\rm m}({\bf x}) \rangle$. However,
the reverse is not true, and it is possible for the gas to contain
only a molecular Bose-Einstein condensate.

The second ingredient is that the symmetries of these two
different phases of the gas are different. In the normal phase the
gas is invariant under the phase transformations $\psi_{\rm
a}({\bf x}) \rightarrow e^{i\theta} \psi_{\rm a}({\bf x})$ and
$\psi_{\rm m}({\bf x}) \rightarrow e^{2i\theta} \psi_{\rm m}({\bf
x})$. The additional factor of two in the transformation of the
molecular annihilation operator follows also from the above
interaction energy and is physically related to the fact that an
molecule consist of two atoms. If the gas contains both an atomic
and a molecular Bose-Einstein condensate (AC+MC) this $U(1)$
symmetry is completely broken, and no residual symmetry exists.
However, if the gas contains only a molecular condensate (MC), a
residual discrete symmetry remains because $\langle \psi_{\rm
m}({\bf x}) \rangle \rightarrow \langle \psi_{\rm m}({\bf x})
\rangle$ for $\theta=\pi$. This phase therefore only breaks the
$U(1)/Z_2$ symmetry spontaneously. As a result there must exist an
Ising-like phase transition between the state with only a MC and
the state with an AC+MC, in which the residual $Z_2$ symmetry is
spontaneously broken. Note that in the case of an atomic Fermi gas
the BCS phase is characterized by a nonzero value of $\langle
\psi_{\rm a}({\bf x}) \psi_{\rm a}({\bf x}) \rangle$. The BCS
phase has therefore the same symmetry as a molecular Bose-Einstein
condensate and only a crossover occurs.

The existence of the Ising transition for the case of the Bose gas
also becomes clear from a consideration of the allowed vortices in
the two limiting cases. The vortices in the state with only a MC
are quantized in integer multiples of an elementary circulation
which is exactly one half the quantum of circulation in the state
with an AC+MC. Consequently there is a
deconfinement-to-confinement transition for a pair of elementary
vortices in the MC with increasing detuning \cite{z2gauge}.
Experiments in the atomic Bose gases thus have the prospect of
observing this phase transition with an essentially topological
character, something which has not so far been possible in
condensed-matter systems.

\begin{figure}[h]
\epsfig{figure=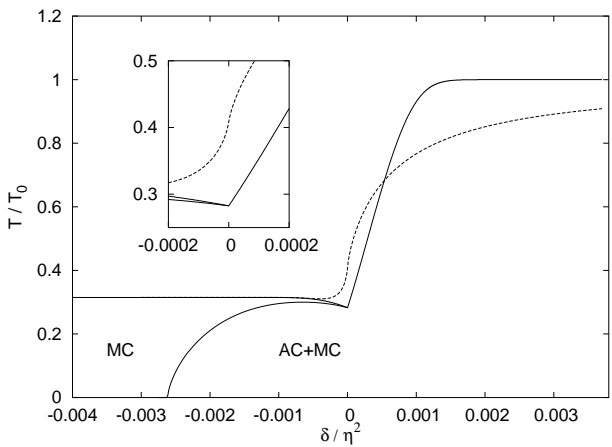,width=8.5cm} \caption{\rm Phase diagram
of a Bose gas with $n=10^{13}$ cm$^{-3}$ near a Feshbach resonance
as a function of detuning $\delta$ and temperature $T$. The solid
lines represents the critical $T$ for the transitions to the
normal state, and the location of the Ising transition between
the phase with only a molecular Bose-Einstein condensate (MC) and
the phase with both an atomic and a molecular Bose-Einstein
condensate (AC+MC). The solid lines do not include the effects of
the finite lifetime of the molecules. The dashed line shows a
calculation of the critical $T$ for the transitions to the normal
state that includes these effects.} \label{endfig1}
\end{figure}

The phase diagram that we expect on the basis of the above
arguments is shown in Fig. \ref{endfig1}. For large positive
$\delta$ the molecular energy lies far above the threshold of the
two-atom continuum and the gas consists essentially completely of
atoms, which Bose-Einstein condense approximately at the ideal gas
critical temperature $T = T_0 \equiv (2\pi\hbar^2/mk_{\rm B})
[n/\zeta(3/2)]^{2/3}$, where $n$ is the total atomic density of
the gas that is assumed to be a constant throughout the phase
diagram. If we lower $\delta$ towards zero, the number of
molecules in the gas increases and consequently the critical $T$
decreases monotonously. For large negative $\delta$, the molecular
energy lies far below the threshold of the two-atom continuum, and
the gas consists solely of molecules that condense at the critical
temperature $T_0/2^{5/3}$. Upon increasing $\delta$ towards zero,
the number of atoms in the gas increases, and the critical $T$ for
Bose-Einstein condensation decreases again. Finally, our arguments
have shown that by increasing $\delta$ at a fixed $T$ below the
critical $T$ for Bose-Einstein condensation of the molecules, a
Ising phase transition to a phase with both an atomic and a
molecular Bose-Einstein condensate occurs. This occurs always at a
negative $\delta$, as we will see next, when we discuss in detail
the quantitative determination of the phase diagram.

{\it Negative detuning.} --- Up to now we have not made a careful
distinction between bare and dressed molecules. However, for a
realistic calculation of the phase diagram this distinction is
very important. The bare molecule corresponds to the bound state
of the Feshbach problem in the absence of a coupling with the
two-atom continuum, whereas the dressed molecule corresponds to
the bound state of the full problem including that coupling. As a
result the energy of the dressed molecule is not equal to the
detuning, but turns out to be given by $\epsilon_{\rm m}=\delta+
\eta^2( \sqrt{1-4\delta/\eta^2}-1 )/2$, where $\eta^2=g^4
m^3/4\pi^2\hbar^6$ and $g=\hbar\sqrt{2\pi a_{\rm bg} \Delta B
\Delta \mu/m}$ determines the coupling between the atoms and the
bare molecules \cite{duine2003a}. The latter can be expressed in
terms of the background scattering length $a_{\rm bg}$, the
magnetic field width $\Delta B$, and the difference between the
magnetic moments of a bare molecule and two atoms $\Delta \mu$,
which are the experimentally know parameters of the Feshbach
resonance. Furthermore, the wave function of the dressed molecule
is not equal to the wave function of the bare molecule, but
contains also amplitude to be in the two-atom continuum. In
detail we find \cite{duine2003b}
\begin{align}
|\chi_{\rm m};{\rm dressed}\rangle = \sqrt{Z}|\chi_{\rm m};{\rm
bare}\rangle +\sum_{\bf k} C_{\bf k}|{\bf k},-{\bf k};{\rm
open}\rangle, \label{open-closed}
\end{align}
where the wave-function renormalization factor $Z$ obeys $1/Z=
1+\eta/2\sqrt{|\epsilon_m|}$. Far from resonance we have that $Z
\simeq 1$ and there is essentially no distinction between the bare
and dressed molecular wave functions. However, close to resonance
$Z$ becomes very small and the dressed molecule actually has
almost no amplitude to be in the bare molecular state.

To take the above physics most easily into account we introduce
creation and annihilation operators for the dressed molecules. The
grand-canonical hamiltonian for the gas then becomes
\begin{align}
\label{hamiltonian} H&=\int dx \psi^{\dagger}_{\rm m}(x)\left[
-\frac{\hbar^2\nabla^2}{4m}+ \epsilon_{\rm m}-2\mu\right]
\psi_{\rm m}(x) \nonumber \\ &+\int dx \psi^{\dagger}_{\rm
a}(x)\left[ -\frac{\hbar^2\nabla^2}{2m}-\mu+\frac{T_{\rm
bg}}{2}\psi^{\dagger}_{\rm a}(x) \psi_{\rm a}(x) \right] \psi_{\rm
a}(x) \nonumber \\ &+\int dx \sqrt{Z} g \left[ \psi^{\dagger}_{\rm
m}(x)\psi_{\rm a}(x)\psi_{\rm a}(x)+\psi^{\dagger}_{\rm
a}(x)\psi^{\dagger}_{\rm a}(x)\psi_{\rm m}(x) \right] \nonumber \\
&+\int dx \frac{T_{\rm mm}}{2}\psi^{\dagger}_{\rm m}(x)
\psi^{\dagger}_{\rm m}(x) \psi_{\rm m}(x)\psi_{\rm m}(x) \nonumber
\\ &+\int dx T_{\rm am} \psi^{\dagger}_{\rm m}(x) \psi^{\dagger}_{\rm
a}(x) \psi_{\rm a}(x) \psi_{\rm m}(x),
\end{align}
where $T_{\rm bg}=4\pi a_{\rm bg}\hbar^2/m$, and $T_{\rm am}$ and
$T_{\rm mm}$ are the T matrices for the scattering of an atom with
a dressed molecule and for the scattering of two dressed
molecules, respectively. Near resonance the latter two T matrices
can actually be expressed in terms of the full atomic scattering
length $a(B)=a_{\rm bg}[1-\Delta B/(B-B_0)]$ of the Feshbach
resonance at magnetic field $B_0$ as we show at the end of this
Letter.

To find the phase diagram for negative $\delta$, we consider the
phase with only a Bose-Einstein condensate of dressed molecules,
and perform a quadratic expansion of the hamiltonian around the
nonzero expectation value $\langle \psi_{\rm m}({\bf x}) \rangle
\equiv \sqrt{n_{\rm mc}}$. For atoms with zero momentum, the
resulting hamiltonian leads to a fluctuation matrix given by
\begin{align}
\left[
\begin{array}{cc}
-\epsilon_{\rm m} + (2T_{\rm am}-T_{\rm mm})n_{\rm mc} & 4 g
\sqrt{Z} \sqrt{n_{\rm mc}} \\ 4 g \sqrt{Z} \sqrt{n_{\rm mc}} &
-\epsilon_{\rm m}+(2T_{\rm am}-T_{\rm mm})n_{\rm mc} \\
\end{array}
\right]. \nonumber
\end{align}
This fluctuation matrix has two positive eigenvalues only if
$\delta$ is sufficiently negative or, more precisely, if
$\epsilon_{\rm m}<-4g\sqrt{Z}\sqrt{n_{\rm mc}}+(2T_{\rm am}-T_{\rm
mm})n_{\rm mc}$. For larger $\delta$ an instability arises, which
physically leads to the formation of an atomic condensate in
addition to the molecular condensate that is already present. This
condition therefore determines the position of the quantum (at
$T=0$) or classical (for $T>0$) Ising phase transition. The
critical $T$ for the Bose-Einstein condensation of the dressed
molecules is of course obtained from the condition $n_{\rm mc} =
0$.

Since both conditions are expressed in terms of the molecular
condensate density $n_{mc}$, we now need to find the equation of
state of the gas to obtain this condensate density as a function
of $\delta$ and $T$. This is complicated by the fact that the
total atomic density of the gas is given by the sum of the density
of atoms and of twice the density of bare molecules, {\em i.e.\/},
not twice the density of the dressed molecules. Using the
techniques derived in Ref.~\onlinecite{RembertReview} the
calculation can nevertheless be performed and we find that in a
good approximation the total density of bare molecules is
\begin{align}
n_{\rm m}&= Z n_{\rm mc} + \frac{1}{V}\sum_{\bf
k}\Bigg(\frac{\epsilon_{\bf k}+2T_{\rm mm}n_{\rm
mc}}{2\hbar\omega_{\bf k}} \frac{1}{e^{\beta \hbar \omega_{\bf
k}}-1} \nonumber \\ &+\frac{\epsilon_{\bf k}+2T_{\rm mm}n_{\rm
mc}-2\hbar\omega_{\bf k}}{4\hbar\omega_{\bf k}}\Bigg),
\end{align}
with $\epsilon_{\bf k}=\hbar^2{\bf k}^2/2m$ and $\hbar\omega_{\bf
k}=\sqrt{\epsilon_{\bf k}^2/4+\epsilon_{\bf k}T_{\rm mm}n_{\rm
mc}}$ the molecular Bogoliubov dispersion. For the atomic density
we find in a similar manner that
\begin{align}
n_{\rm a}&= \frac{1}{V}\sum_{\bf k}\Bigg(\frac{2\epsilon_{\bf k}-
\epsilon_{\rm m}+(2T_{\rm am}-T_{\rm mm})n_{\rm mc}}
{2\hbar\omega_{\bf k}} \frac{1}{e^{\beta \hbar \omega_{\bf k}}-1}
\nonumber \\ &+\frac{2\epsilon_{\bf k}- \epsilon_{\rm m}+(2T_{\rm
am}-T_{\rm mm})n_{\rm mc} -\hbar\omega_{\bf k}}{4\hbar\omega_{\bf
k}}\Bigg),
\end{align}
where the dispersion for the atoms obeys $\hbar\omega_{\bf
k}=\sqrt{[\epsilon_{\bf k}- \epsilon_{\rm m}/2+(T_{\rm am}-T_{\rm
mm}/2)n_{\rm mc}]^2 - 4g^2Zn_{\rm mc}}$. Note that this dispersion
in general has a gap, but becomes gapless exactly at the critical
condition for the Ising transition, as expected.

{\it Positive detuning.} --- At positive $\delta$ no truly stable
molecular state exists and the previous approach in principle does
not apply. That approach was based on the fact that for negative
$\delta$, the density of states for the bare molecules contained a
delta function at the energy $\epsilon_{\rm m}$ and with strength
$Z$, that corresponded to the dressed molecules. For positive
$\delta$ the density of states consists, however, only of a single
broad peak \cite{duine2003a}. Approximating that broad peak by a
delta function at the energy where the density of states has a
maximum, we can again use the hamiltonian in
Eq.~(\ref{hamiltonian}), but now with
\begin{align}
\epsilon_{\rm m}
=\frac{1}{3}\left(\delta-\frac{\eta^2}{2}+\sqrt{\frac{\eta^4}{4}
-\eta^2\delta+4\delta^2}\right)
=\frac{\delta^2}{\eta^2}+O\left(\delta^3\right)
\end{align}
and $Z=1$. In this approximation we are neglecting the finite
lifetime of the bare molecules, that is due to the fact that these
molecules can now decay into the two-atom continuum. At the end of
this Letter we, however, also discuss a calculation that includes
these finite lifetime effects exactly \cite{RembertReview}. The
result of that calculation is indicated by the dashed line in
Fig.~1, which shows that the above approximation is indeed rather
accurate. Since for positive $\delta$ we only need to determine
the critical $T$ for Bose-Einstein condensation of the atoms and
molecules, we here consider only the normal phase of the gas. At
the critical temperature $T_{\rm c}$ the atomic density is
therefore just $n_{\rm a}= n (T_{\rm c}/T_0)^{3/2}$, whereas the
density of molecules equals
\begin{align}
n_{\rm m}=2\sqrt{2} n \left(\frac{T_{\rm c}}{T_0}\right)^{3/2}
\frac{g_{3/2}(e^{-\beta\epsilon_{\rm m}})}{\zeta(3/2)}~,
\end{align}
where $g_{3/2}(z)$ is the usual Bose function. The desired
critical $T$ now follows from $n = 2n_{\rm m} + n_{\rm a}$.

\begin{figure}[h]
\epsfig{figure=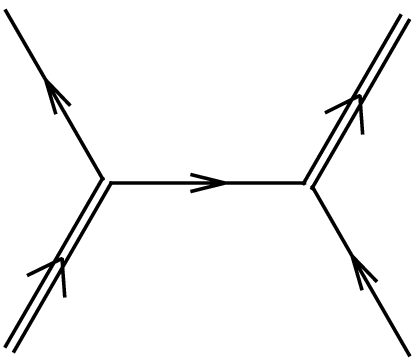,width=2.5cm}
\epsfig{figure=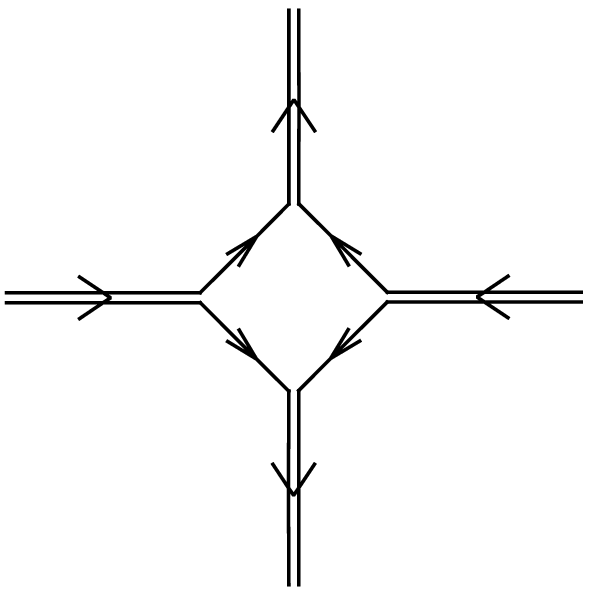,width=2.5cm} \caption{\rm Feynman
diagrams that determine a) the interaction between an atom and a
dressed molecule and b) the interaction between two dressed
molecules. \label{interactions}}
\end{figure}

{\it Discussion and conclusions.} --- Up to now we have left the
molecule-molecule interaction and the atom-molecule interaction
unspecified. Near resonance they are however in first
approximation determined by the Feynman diagrams given in
Fig.~\ref{interactions}. Evaluating these diagrams at zero
external momenta and frequencies, we find that $T_{\rm am}=8 g^2
Z/|\epsilon_{\rm m}|$. Close to resonance we have that $Z = 4 \pi
\hbar^4/g^2 m^2 a$ and $|\epsilon_{m}|=\hbar^2/ma^2$. We thus
conclude that the scattering length for this process is
proportional to the atom-atom scattering length as $a_{\rm
am}=32a/3$. Using a similar procedure, the scattering length for
the molecule-molecule interaction can be shown to be $a_{\rm mm}=4
a$. In the case of an atomic Fermi gas we find in the same manner
that $a_{\rm am}= 8a/3$ and $a_{\rm mm}=a$. This may be compared
to similar results obtained recently by Petrov {\it et al.} for
this case \cite{PetrovArchive}.

In the calculation of the phase diagram presented in
Fig.~\ref{endfig1} we have actually not included the mean-field
effects due to the atom-molecule and molecule-molecule
interactions. The reason is that these mean-field effects, if we
also take into account the unitarity limit of the T matrix, are
estimated to lead only to relatively small shifts. Qualitatively,
the most important effect is that the Ising transition is, near
$T=0$, shifted to somewhat higher values of $\delta$. In
particular, the transition temperatures for Bose-Einstein
condensation are hardly affected. As mentioned previously, we have
performed a calculation of the critical $T$ for Bose-Einstein
condensation as a function of $\delta$, that includes both the
effects of the finite lifetime of the molecule at positive
$\delta$, as well as the rogue-dissociation process \cite{mackie}
for negative $\delta$. The results are presented by the dashed
line in Fig.~\ref{endfig1}. Note that for negative $\delta$ these
effects are nonnegligible only in a small region close to
resonance, due to the fact that $Z$ becomes very small, which
implies that there is a rather large spectral weight in the
density of states at positive energies. In addition,
$\epsilon_{\rm m}$ becomes very small, which implies that the
positive energy states also get thermally populated. For positive
$\delta$ the results are affected more, because of the fact that
the finite lifetime of the molecules leads to a substantial
broadening of the density of states, which considerably affects
the total number of bare molecules in the gas. In future work we
intend to determine also the effects of rogue dissociation on the
critical line of the Ising transition. We also want to explore the
possibility that sufficiently close to resonance, the continuous
Ising transition is actually preempted by a first-order phase
transition by such effects.

Field theories for critical fluctuations near the Ising
transitions at $T=0$ and $T>0$ can be developed by standard
methods, using that in the states with $\langle \psi_{\rm m}
\rangle \neq 0$ the Ising order parameter $\phi \propto
i(\psi_{\rm a} - \psi_{\rm a}^{\dagger})$. These field theories
have the familiar $\phi^4$ form, except that at $T=0$ there is
also a marginal coupling to dynamic density fluctuations of the
superfluid which apparently drives the transition weakly
first-order \cite{frey} (the $T>0$ transition can remain second
order). The critical precursors of the transition lead to a large
density of states for collective low energy excitations, which
will strongly damp single-particle excitations. We speculate that
observation of such enhanced sources of dissipation may serve, in
addition to direct observation of the MC and AC+MC, as an
experimental signature of the Ising transition.

While the writing of this paper was being completed, we became
aware of the work of Radzihovsky {\it et al.} \cite{leo}, who also
point out the existence of the Ising phase transition, but use the
Timmermans model \cite{timmermans1999b} to determine the phase
diagram. This work is supported by the Stichting voor Fundamenteel
Onderzoek der Materie (FOM), the Nederlandse Organisatie voor
Wetenschaplijk Onderzoek (NWO), and the US National Science
Foundation grant DMR-0098226.

\bibliographystyle{apsrev}

\begin{thebibliography}{104}
\bibitem{z2gauge} In three dimensions at nonzero temperatures, a
                  duality mapping shows that this is the confinement
                  transition in a $Z_2$ gauge theory that is dual to the
                  Ising model.
\end{thebibliography}

\end{document}